\documentclass[letterpaper, 12pt, titlepage, reqno]{article}

\usepackage[letterpaper, portrait, margin=2.5cm]{geometry}

 \usepackage[super]{natbib}
 \usepackage{amsmath, amssymb, latexsym}

\usepackage{graphicx}% Include figure files
\usepackage{epsfig}
\usepackage{epstopdf}
\usepackage{caption}

\newcommand{\bs}{\mbox{${\bf s}$}}

\newcommand{\bx}{\mbox{${\bf x}$}}

\newcommand{\bD}{\mbox{${\bf D}$}}

\newcommand{\bI}{\mbox{${\bf I}$}}

\newcommand{\bthetahpr}{\mbox{$\hat{\mbox{\boldmath $\bf \theta$}}
            ^{\raise-.5ex\hbox{$\scriptstyle\prime$}}$}}
\newcommand{\bphihpr}{\mbox{$\hat{\mbox{\boldmath $\bf \phi$}}
            ^{\raise-.5ex\hbox{$\scriptstyle\prime$}}$}}
\newcommand{\bThetahpr}{\mbox{$\hat{\mbox{\boldmath $\bf \Theta$}}
            ^{\raise-.5ex\hbox{$\scriptstyle\prime$}}$}}
\newcommand{\bPhihpr}{\mbox{$\hat{\mbox{\boldmath $\bf \Phi$}}
            ^{\raise-.5ex\hbox{$\scriptstyle\prime$}}$}}

\newcommand{\subearth}{\raise.15ex\hbox{$\scriptstyle\oplus$}}

\newcommand{\subspace}{\raise.25ex\hbox{$\scriptscriptstyle\bigcirc$}}

\newcommand{\bdel}{\mbox{\boldmath $\nabla$}}

\newcommand{\tdot}{\,.\hspace{-0.98 mm}\raise.6ex\hbox{.}
                   \hspace{-0.98 mm}\raise1.2ex\hbox{.}\,}

\newcommand{\allspace}{\raise.25ex\hbox{$\scriptstyle\bigcirc$}}

\newcommand{\eq}{\begin{equation}}
\newcommand{\en}{\end{equation}}
\newcommand{\eqa}{\begin{eqnarray}}
\newcommand{\ena}{\end{eqnarray}}

\renewcommand{\epsilon}{\varepsilon}

\newcommand{\rmd}{\, \mathrm{d}}

\begin{document}

 \begin{titlepage}
 \begin{center}

\textbf{Anelastic sensitivity kernels with parsimonious storage for adjoint tomography and full waveform inversion}\\

 \vspace{10ex}

Dimitri Komatitsch$^1$, Zhinan Xie$^{2,1}$, Ebru Bozda{\u{g}}$^{3}$, Elliott Sales de Andrade$^{4}$,
Daniel Peter$^{5}$, Qinya Liu$^{4}$, and Jeroen Tromp$^{6}$\\

 \vspace{5ex}

 ~$^1$LMA, CNRS UPR 7051, Aix-Marseille University, Centrale Marseille, 13453 Marseille cedex 13, France\\
  E-mail: komatitsch@lma.cnrs-mrs.fr\\
 ~$^2$Institute of Engineering Mechanics, China Earthquake Administration, Harbin 150080, China\\
 ~$^3$G\'eoazur, University of Nice Sophia Antipolis, Valbonne, France\\
 ~$^4$Department of Physics and Department of Earth Sciences, University of Toronto,\\ Toronto, Ontario, Canada\\
 ~$^5$Extreme Computing Research Center, King Abdullah University of Science and Technology (KAUST),\\ Thuwal, Saudi Arabia\\
 ~$^6$Department of Geosciences and Program in Applied \& Computational Mathematics,\\ Princeton University, New Jersey, USA.
 \end{center}
 \end{titlepage}

\begin{abstract}
We introduce a technique to compute exact anelastic sensitivity kernels in the time domain using parsimonious disk storage. The method is based on a reordering of the time loop of time-domain forward/adjoint wave propagation solvers combined with the use of a memory buffer. It avoids instabilities that occur when time-reversing dissipative wave propagation simulations. The total number of required time steps is unchanged compared to usual acoustic or elastic approaches. The cost is reduced by a factor of 4/3 compared to the case in which anelasticity is partially accounted for by accommodating the effects of physical dispersion. We validate our technique by performing a test in which we compare the $K_\alpha$ sensitivity kernel to the exact kernel obtained by saving the entire forward calculation. This benchmark confirms that our approach is also exact. We illustrate the importance of including full attenuation in the calculation of sensitivity kernels by showing significant differences with physical-dispersion-only kernels.
\end{abstract}

\noindent \begin{center}{\textbf{*** This manuscript is now published as a paper in the Geophysical Journal International, 2016. ***}}\end{center}

\section{Introduction}
\label{introduction}

Efficient numerical methods for simulating the propagation of acoustic, elastic, or anelastic waves in the time domain are widely available,
for instance based on finite-difference methods~\citep[see e.g.,][for a review]{ViOp09},
spectral-element methods~\citep[e.g.,][]{KoVi98,VaCaSaKoVi99,KoTr99,KoTr02a}, or standard finite-element methods~\citep[e.g.,][]{KaFaKuStBiGh13}.
Nowadays, these techniques are heavily used for imaging based on full waveform inversion (FWI)
or adjoint tomography~\citep[e.g.,][]{TrTaLi05,Plessix_2006_RAS,TrKoLi08,ViOp09,Fic10,MoChKoWa15}.
FWI involves fitting band-pass filtered versions of observed seismograms by minimizing least-squared differences between observed and synthetic seismograms.
Adjoint tomography generalizes FWI by considering arbitrary measures of misfit,
e.g., cross-correlation traveltimes, multi-taper phase and amplitude anomalies, or instantaneous phase measurements.

In the context of imaging, it is useful to resort to the concept of sensitivity
kernels~\citep[e.g.,][]{Tarantola_1986_SNL,Tar87,Tar88,TrTaLi05,TrKoLi08,LiTr08,Fic10,FiVa14}.
Let~$\bs$ denote the forward displacement wavefield and~$\bs^\dagger$ the adjoint wavefield.
In an isotropic Earth model, the kernels~$K_\kappa$ and~$K_\mu$ represent Fr\'echet derivatives with respect to relative
bulk and shear moduli perturbations, respectively. These kernels are given by~\citep[e.g.,][]{TrKoLi08}
\begin{equation}
K_\kappa(\bx) = - \int_0^T \kappa(\bx)\,[\bdel \cdot \bs^\dagger(\bx,T-t)][\bdel \cdot \bs(\bx,t)]\rmd t \, ,
\label{definition_of_Kkappa}
\end{equation}
\begin{equation}
K_\mu(\bx) = - \int_0^T 2 \mu(\bx)\,\bD^\dagger(\bx,T-t):\bD(\bx,t)\rmd t \, ,
\label{definition_of_Kmu}
\end{equation}
where
\begin{equation}
\bD = \frac{1}{2} [\bdel\bs + (\bdel\bs)^\mathrm{T}] - \frac{1}{3} (\bdel \cdot \bs)\,\bI
\end{equation}
and
\begin{equation}
\bD^\dagger = \frac{1}{2} [\bdel\bs^\dagger + (\bdel\bs^\dagger)^\mathrm{T}] - \frac{1}{3} (\bdel \cdot \bs^\dagger)\,\bI
\end{equation}
denote the traceless strain deviator and its adjoint, respectively,~$\bx$ is the position vector,
and~$\kappa$ and~$\mu$ are the bulk and shear moduli, respectively.
Their expression remains valid for elastic perturbations superimposed on an anelastic Earth model \citep{LiTr08}
if the regular and adjoint wavefields are computed in that anelastic reference model.
In practical applications, it is often useful to define compressional and shear wavespeed sensitivity kernels, namely \citep{TrTaLi05,TrKoLi08}
\begin{equation}
K_\alpha = 2 \left( \frac{\kappa + \frac{4}{3} \mu}{\kappa} \right) K_\kappa
\label{definition_of_Kalpha}
\end{equation}
and
\begin{equation}
K_\beta = 2 \left( K_\mu - \frac{4}{3} \frac{\mu}{\kappa} K_\kappa \right) \, .
\label{definition_of_Kbeta}
\end{equation}
In order to perform the convolution involved in the calculation of the kernels~(\ref{definition_of_Kkappa}) and~(\ref{definition_of_Kmu}),
simultaneous access to the forward wavefield~$\bs$ at time~$t$ and the adjoint wavefield~$\bs^\dagger$ at time~$T-t$ is required
(or conversely, since~$\int_0^T f(t)\,f^\dagger(T-t) \rmd t = \int_0^T f(T-t) \,f^\dagger(t) \rmd t$ when convolving two functions~$f$ and~$f^\dagger$).
Carrying out forward and adjoint simulations simultaneously is insufficient,
because in that case both wavefields are only available at a given time~$t$.
In addition, to calculate the adjoint wavefield one must prescribe the adjoint source,
and that source is computed based on measurements between observed and simulated seismograms,
i.e., it can only be constructed after the completion of a forward simulation.

A straightforward solution to this dilemma is to store the entire forward simulation to disk and then read it back in reverse order
during the adjoint simulation. For 1D or 2D models this is feasible~\citep[e.g.,][]{PaKaNe16}, but in 3D at short periods
without lossy compression or significant spatial or temporal sub-sampling~\citep{FiKeIg09,SuFu13,RuHaPuGu14,CyShWi15}
the required amount of disk storage is currently prohibitive.
It is worth mentioning that this situation will likely change in the future, but not any time soon.
In addition, heavy I/O involved in reading back the forward wavefield can significantly slow down the simulation~\citep{YuWeLiZhLi14}.

For nondissipative acoustic or elastic media,
a standard solution for large 3D problems is to perform three simulations per source~\citep{TrKoLi08,PeKoLuMaLeCaLeMaLiBlNiBaTr11}.
One performs the forward calculation twice: once to compute the adjoint sources and once again in reverse time
simultaneously with the adjoint simulation performed in forward time to correlate the two fields and sum their interaction on the fly over all time steps.
Thus, one only needs a small amount of disk storage to store the last time
step of the forward run, which is then used as an initial condition to recalculate the forward wavefield backwards in time with a negative time step.
Based on this strategy, one has simultaneous access to the adjoint wavefield at time~$t$ and the forward wavefield at time~$T-t$,
which is what is required to perform the convolution involved in the construction of the kernels~(\ref{definition_of_Kkappa}) and~(\ref{definition_of_Kmu}).
In the anelastic case, as shown by \cite{Tar88}, the wave equation is no longer self adjoint, leading to exponential growth of energy in the
adjoint equation  (i.e., `anti attenuation').
However, in the calculation of sensitivity kernels this does not matter, because this process involves the {\em time-reversed\/} adjoint wave equation rather
than the adjoint wave equation.
In this respect, it is worth mentioning that \cite{LiTr08} have a time-reversed definition of the adjoint state compared to the classical one
of e.g. \cite{Tar88} and \cite{ViOp09}, i.e. they use the time-reversed equation~(35) of \cite{Plessix_2006_RAS} instead of his equation~(32).
Thus `anti attenuation' in forward time becomes attenuation in reverse time,
and the time-reversed adjoint wave equation is identical to the classical forward wave equation~\citep{Tar88,LiTr08,FiBuIg06}.
Consequently, both the forward and the adjoint wavefields are attenuated during the calculation of the kernel.
This makes physical sense, because as the distance between two seismic stations increases,
one expects the sensitivity kernel to become weaker and weaker as a result of dissipation.
Unfortunately, in the presence of attenuation the process of reconstructing the forward wavefield backwards in time based on the final snapshot
and a negative time step is numerically highly unstable~\citep[e.g.,][]{LiTr06,LiTr08,KoSc11,AmBrGaWa13}.
This results in numerical instabilities during the calculation of anelastic sensitivity kernels,
not because of the adjoint run, but because of the forward run that needs to be performed backwards in time.

An approximate solution to this conundrum involves modifying the wave equation to introduce filtering or other stabilizing terms,
or incorporating only certain aspects of anelasticity.
\cite{AmBrGaWa13} introduced a promising regularized time-reversal imaging technique which corrects attenuation effects to first order.
However, their approach involves significant filtering, thereby affecting the quality of the resulting signals, in particular at high frequencies,
and it is currently limited to simple models of weak attenuation in homogeneous media.
In particular, the approach cannot handle models comprised of standard linear solids.
\cite{Zhu14} and~\cite{ZhHaBi14} introduce partial support for attenuation in reverse-time migration
by separating amplitude attenuation and phase dispersion operators~\citep{VaRoUl93}.
They construct attenuation- and dispersion-compensated operators by reversing the sign of the attenuation
operator and leaving the sign of the dispersion operator unchanged; they then design a low-pass filter for these operators
to stabilize the numerical procedure and avoid amplifying high-frequency noise that would trigger instabilities.
Changing the sign of one of the terms in the backward integrations to stabilize the calculations is also classically done
in so-called Back-and-Forth Nudging algorithms~\citep{AuBlNo11,AuBaBl13}.

An alternative solution consists of resorting to so-called `partial checkpointing' or `optimal checkpointing',
i.e., using partial storage to disk, but, realizing that storage size limitations or slowdown related to disk storage are important issues~\citep{YuWeLiZhLi14},
defining an optimized sequence in which the forward and adjoint time steps are performed,
essentially trading storage requirements for longer computation times.
This elegant idea was introduced by~\cite{ReLeGr98} and~\cite{GrWa00}.
\cite{ReLeGr98} used a recursive strategy to compute the optimized order in which the simulation steps need to be performed
and showed that when the schedule is optimized the storage and computational times grow at most logarithmically,
and~\cite{GrWa00} defined an algorithm called `Revolve' that is provably optimal to reduce storage requirements.
\cite{Cha01} used it for the meteorological model `Meso-NH' and proposed several variants depending on user preference
between CPU time and memory optimization.
\cite{AkBiBiEpFeGhKiLoOhTuUr03},~\cite{Sym07}, and~\cite{AnTaWa12} resorted to it for FWI as well as reverse-time migration.
\cite{HiWaSt06} applied it to the instationary Navier-Stokes equations, and more recently~\cite{SpCoKa14}
included it in a jet-engine noise reduction simulation code.
A limitation of the original `Revolve' algorithm of~\cite{GrWa00}
is that it requires {\it a priori} knowledge of the total number of time steps to be performed,
making it incompatible with adaptive time stepping.
\cite{WaMoIa09} have addressed that issue by introducing a dynamic checkpointing algorithm
that is applicable even when the total number of time steps is {\it a priori} unknown.

\begin{figure}
\centerline{\includegraphics[width=0.65\linewidth]{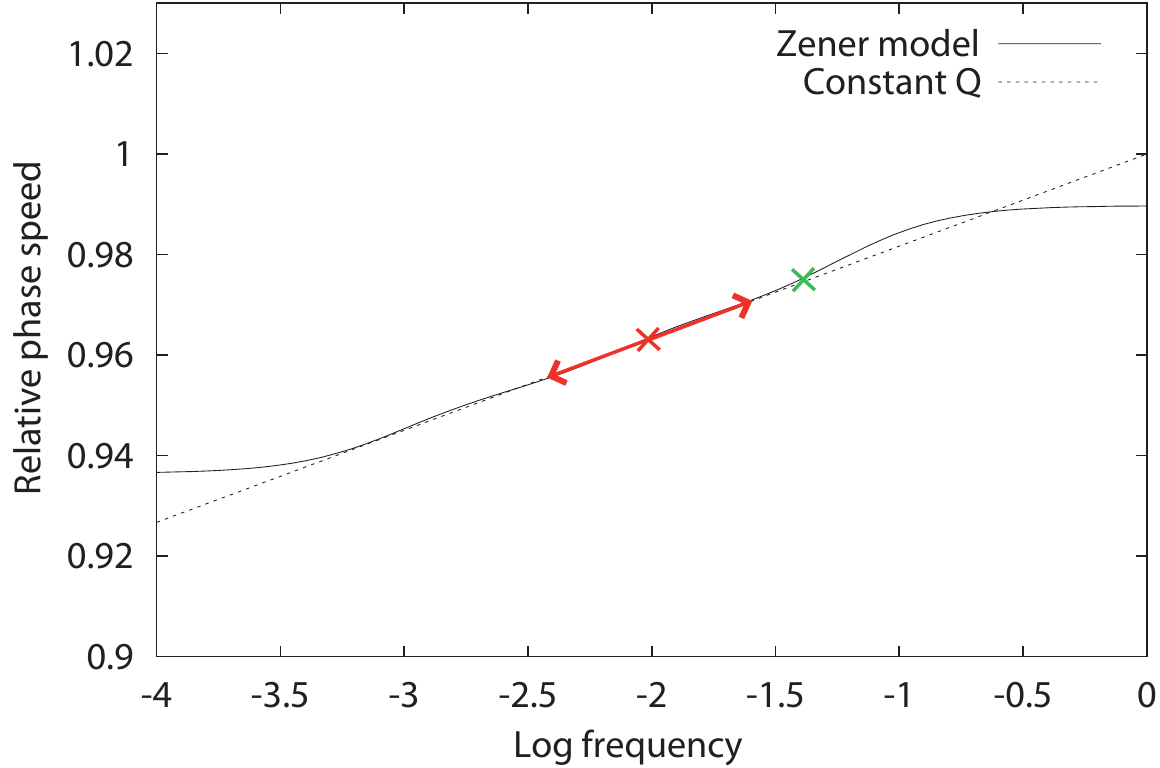}}
\caption{\small Taking into account only the effects of physical dispersion in anelastic simulations,
i.e., its effect on phase but not on amplitudes (no dissipation), amounts to performing an elastic
simulation in a wavespeed model shifted to the dominant frequency of the source (red cross) relative to the frequency of the reference
model (green cross), based on equation~(\ref{eq:PD}).
The logarithmic phase speed in a strictly constant-$Q$ absorption-band model is represented by the dotted line.
In practice, in time-domain simulations a constant-$Q$ model is approximated by a small number of standard linear solids (usually Zener solids) in
parallel~\citep[e.g.,][]{Car14,BlKoChLoXi16}, which approximate a constant~$Q$ inside an absorption band of interest (solid line).}
\label{taking_into_account_average_physical_dispersion}
\end{figure}

Another ---more precise, but also more expensive--- approximate solution is to accommodate the effects of physical dispersion
induced by attenuation, but not the effects of dissipation.
To appreciate this, consider the frequency-dependent shear modulus in a constant-Q absorption-band solid, namely~\citep[e.g.,][]{LiAnKa76,DaTr98,Car14}
\begin{equation}
 \frac{\mu(\omega)}{\mu(\omega_0)}=1+\frac{2}{\pi Q_\mu}\,\ln\left(\frac{\omega}{\omega_0}\right)+\frac{i}{Q_\mu} \, .
 \label{eq:shear}
\end{equation}
Here~$\omega$ denotes the angular frequency of interest, $\omega_0$ a chosen reference frequency, and~$Q_\mu$ a frequency-independent shear quality factor.
The effects of physical dispersion are captured by the logarithmic term, $2/(\pi Q_\mu)\ln(\omega/\omega_0)$\,,
and the effects of dissipation by the complex part of the modulus, $1/Q_\mu$\,.
Based on these observations, partial support for attenuation is accommodated by performing
a total of four simulations per source~\citep[e.g.,][]{ZhLiTr11,ZhBoPeTr12},
instead of three in the nondissipative case~\citep{TrKoLi08,PeKoLuMaLeCaLeMaLiBlNiBaTr11}.
In a first stage, one runs a forward simulation twice in the forward direction, once with full attenuation and once with physical dispersion only.
The second calculation is therefore a purely elastic calculation, but for a wavespeed model that is shifted to the dominant frequency of the source,
$\omega_s$\,, based on the correction
\begin{equation}
 \mu(\omega_s)=\mu(\omega_0)[1+2/(\pi Q_\mu)\ln(\omega_s/\omega_0)] \, ,
 \label{eq:PD}
\end{equation}
as illustrated in Figure~\ref{taking_into_account_average_physical_dispersion}.
The first forward run with full attenuation is used to make the measurements needed in the construction of the adjoint sources for
the third run, and the second forward run with physical dispersion only is used to compute and store
the final time step to be able to time-reverse that calculation in the fourth run.
The third and fourth runs are carried out simultaneously to calculate the kernel,
and both these runs are purely elastic and use a wavespeed model that is shifted to the dominant frequency of the source.
The fourth run is thus stable, because it involves time reversal of an elastic simulation.
Note, however, that the measurements that are assimilated in the third run are based on synthetics computed with full attenuation.

It is important to recognize that the significance of anelastic effects on sensitivity kernels very much depends on the type of
measurement one chooses to make. For example, as demonstrated by~\cite{TrTaLi05},
the sensitivity kernel for a cross-correlation traveltime measurement is identical to the so-called `banana-donut' kernel first introduced by~\cite{DaHuNo00}.
Such kernels may be calculated based on ray theory,
and the related expressions are largely unaffected by attenuation.
Physically, this reflects the fact that traveltimes are affected by wavespeed, and only very marginally by dissipation.
More generally, as long as one focuses on measuring phase,
e.g., frequency-dependent traveltime or instantaneous phase,
the corresponding kernels are largely unaffected by attenuation~\citep{ZhLiTr11}.
However, for inversions involving amplitude measurements, including FWI, attenuation plays a critical role.
More generally, attenuation is important on a global scale~\citep[e.g.,][]{RuZh10,RuZh12},
in exploration geophysics~\citep[e.g.,][]{KuPrKoBo13,GrScFoBo14},
and in near-surface geophysics or for site effects in poorly consolidated sediments~\citep[e.g.,][]{AsAkBiGh07,AsKaKaLiKu12}.
We demonstrate in this paper that global simulations of surface waves at periods less than 40--60~s must accommodate the full effects of attenuation.

To some extent, the accuracy of the gradient is not that critical in the early stages of an inversion,
because as part of the iterative inversions scheme,
e.g., a non-linear conjugate gradient method or a L-BFGS quasi-Newton method,
the raw gradient is generally smoothed and preconditioned~\citep[e.g.,][]{ShUl90,Fel92,Shi06}.
Nevertheless, as the inversion proceeds and the frequency content of the seismograms is increased,
details in the gradient matter, and its accurate calculation becomes highly relevant.

\section{Parsimonious storage technique}

\begin{figure}
\centerline{\includegraphics[angle=-90,width=0.75\linewidth]{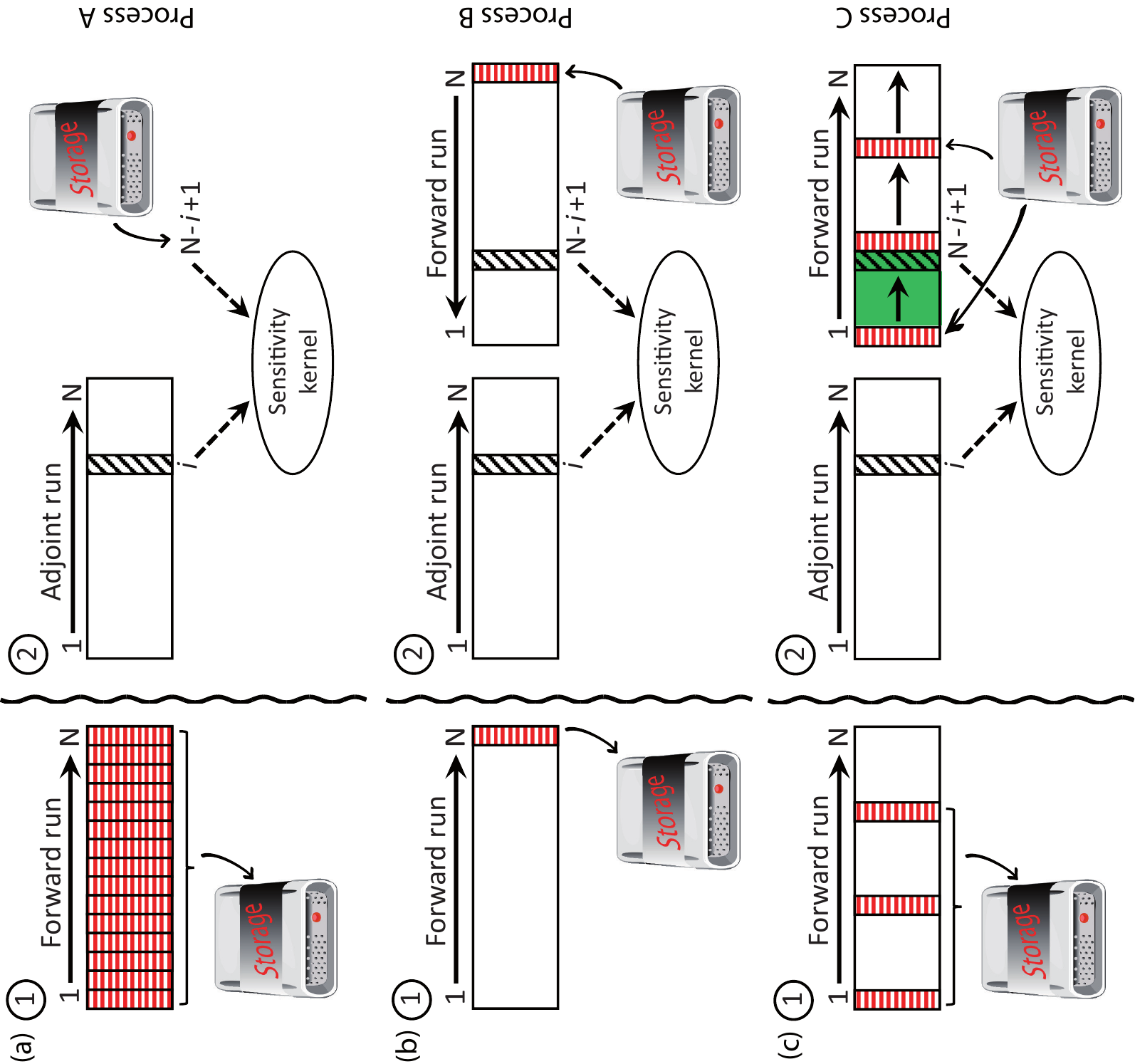}}
\caption{\small For time-domain simulations consisting of~$N$ time steps numbered from 1 to~$N$,
the contribution to adjoint-based sensitivity kernels at time step~$i$
is obtained by combining information coming from time step~$i$ of the adjoint
run with information coming from time step~$N-i+1$ of the forward run.
For low to moderate frequencies and/or small to moderate model sizes, one can consider storing the entire forward run to disk (a, red dashed slices),
reading it back from disk in reverse order while computing the adjoint wavefield.
However, for high frequencies and/or large model sizes
the amount of storage needed is currently unaffordable.
Another classical approach is to perform the forward run first and store its final time step to disk,
and in a second stage perform the adjoint run while simultaneously redoing the forward run backwards,
reversing time and starting from the stored final time step~(b).
In the acoustic or elastic cases, that process is stable, but not in the anelastic case.
However, on computers that have a significant amount of memory per compute node,
a third process can be designed, which is stable even in the presence of energy loss~(c). During the first stage one saves
checkpointing/restart files to disk every few hundred or thousand time steps. During the second stage one still performs two runs simultaneously,
but instead of performing the forward run backward from the stored final time step one performs it in chunks
in reverse order but in the {\em forward} direction, in each case starting from the previous restart file read back from disk
and storing that sub-part of the run in memory (the green region, stored in a memory buffer).
Since the run is conducted forward rather than backward in time, this process is always stable, even in the presence of attenuation.}
\label{how_we_can_undo_attenuation}
\end{figure}

In view of eqns.~(\ref{definition_of_Kkappa}) and~(\ref{definition_of_Kmu}),
for simulations in the time domain consisting of~$N$ time steps numbered from~1 to~$N$,
the contribution to adjoint-based sensitivity kernels at time step~$i$
is obtained by combining information coming from time step~$i$ of the adjoint
run simultaneously with information coming from time step~$N-i+1$ of the forward run.
Two classical approaches can be used to facilitate this, namely,
at low to moderate frequencies and/or small to moderate model sizes
one can consider storing the entire forward run to disk (Process~A; see Figure~\ref{how_we_can_undo_attenuation}a),
and reading it back from disk in reverse order during the adjoint run.
Lossless or lossy compression~\citep{FiKeIg09,SuFu13,RuHaPuGu14,CyShWi15} or spatial or temporal sub-sampling~\citep{SuFu13} can be helpful in this context.
However, the required amount of storage is currently unaffordable and will remain so for many years to come, and, in addition, heavy I/O
will significantly slow down the simulation code~\citep{YuWeLiZhLi14}.

Another classical approach~\citep[e.g.,][]{LiTr06,LiTr08,TrKoLi08} is to first perform the forward run and store its final time step to disk,
and in a second stage perform the adjoint run in the forward direction
while simultaneously redoing the forward run backwards, reversing time and starting from the
final time step (Process~B; see Figure~\ref{how_we_can_undo_attenuation}b).
In the acoustic or elastic case, i.e., when total energy is conserved, this process is numerically stable and its only two drawbacks
are that the compute time increases by a factor of 3/2 because the forward run needs to be performed twice,
and that the required memory size increases by a factor of two
because during the second stage two runs need to be performed simultaneously in memory.
Let us note that the first drawback is not that serious compared to Process~A (Figure~\ref{how_we_can_undo_attenuation}a),
because heavy I/O slows the latter down down considerably.
Unfortunately, as mentioned above, Process~B cannot be used ---without heavy filtering
and resulting significant loss of accuracy~\citep[e.g.,][]{AmBrGaWa13}--- in the anelastic case or in the presence of any kind of energy loss,
because time-reversing energy decay is unstable from a numerical point of view~\citep[e.g.,][and references therein]{LiTr06,LiTr08,KoSc11,AmBrGaWa13}.
The reason is that while amplifying the fields to restore energy when going backwards,
numerical schemes will also amplify numerical noise and thus very quickly become unstable.

Even if full attenuation cannot be taken into account in Process~B, it can be partially accommodated
by performing two runs instead of one during Stage~1, as we will see in more details in Section~\ref{SectionImportance}:
one with full attenuation to make measurements to be used in the calculation
of the adjoint sources for Stage~2, and another one with physical dispersion only in order to compute and store the final time step
and be able to time-reverse that calculation in Stage~2.
In the literature it is often mentioned~\citep[e.g.,][]{LiTr06,LiTr08} that
during Stage~2 one needs to reverse time and perform the forward run backwards,
but, more precisely, the only requirement is to have access to time step~$N-i+1$ of the forward run, regardless of whether
it is computed backward or forward in time.
Thus, on computers that have a significant amount of memory
---which is always the case on modern compute clusters--- a third process can be designed, which is stable
even in the presence of energy loss (Process~C; see Figure~\ref{how_we_can_undo_attenuation}c).
In this approach, during the first stage one saves
a small number of evenly-spaced checkpointing/restart files of the three components of the displacement field to disk,
typically one every few hundred or thousand time steps;
during the second stage one still performs two simulations simultaneously,
one adjoint run and one forward run, but instead of performing the forward run {\em backward} from the stored final time step one performs it in chunks,
in reverse order, but in the {\em forward} direction inside each chunk.
In each instance one starts from the previous restart file of displacement read back from disk,
storing only that sub-part of the run in memory.
Since the run is conducted forward rather than backward in time, this process is always stable, even in the presence of attenuation.
It is also exact, since no filtering is involved.
Process~C is computationally meaningful compared to Process~A only if the number of time steps between two checkpoints is sufficiently large,
say a few hundred to a few thousand, i.e., if the total memory available as a storage buffer is large enough.
Fortunately, in practice that is almost always the case on modern compute clusters.

Compared to the more involved `Revolve' algorithm of~\cite{GrWa00} discussed in the introduction,
which is provably optimal in terms of minimizing the number of time steps to store in memory~\citep[see also, e.g.,][]{HiWaSt06},
our choice is different and rather makes optimal use of the entire computer memory,
i.e., it maximizes memory usage instead of trying to minimize it.
The rationale for this is that monitoring of typical large wave propagation
simulations, for instance in seismology or in the oil industry,
shows that ---considering the large compute clusters or supercomputers
which are nowadays readily available--- users use only a small portion of the memory available per compute node, typically between 5\% and 30\%,
because they generally harness a relatively large number of processor cores to keep the calculation relatively fast.
Thus, leaving 5\% for the operating system of the machine, between 65\% and 90\% of the total memory is available and can be used
as a memory buffer. This enables one to store at least hundreds of time steps of the displacement vector, sometimes even a few thousand.
The number of time steps that can be stored is readily computed in an exact fashion once and for all before the time loop
by dividing the size of the available free computer memory by the (constant) size of the array that contains the three components of the
displacement vector at a given time step. Note that sensitivity kernel calculations
often require the strain~\citep[e.g.,][as well as eqns.~\ref{definition_of_Kkappa} and~\ref{definition_of_Kmu}]{TrKoLi08,LiTr08},
but to reduce storage to disk, which is both disk-space consuming and slow,
we usually recompute the strain from the stored displacement instead of storing it.

In terms of implementation, adding this approach to an existing code is easy because it consists mainly of
restructuring the time loop and implementing a simple memory buffer system.
Note that the cost does not increase compared to Process~B, because we perform the same total number of operations, simply in a different order.
In fact, if attenuation is partially taken into account in Process~B, four runs are needed instead of three, as mentioned above,
and in such a case Process~C is cheaper by a factor~4/3.
The writing of checkpointing files to disk during Stage~1 of the algorithm
may be non-blocking (if technically feasible on the file system), thereby allowing for overlapping
of disk writes with calculations, because these restart files are reused much later in the algorithm, during Stage~2.
The reason why a memory buffer is needed during Stage~2 is that in order to gain access to time step~$N-i+1$ of the forward run,
this buffer will be filled in forward order from the previous restart file, but will then be accessed backward,
i.e., in reverse order from its end, a policy often called `Last In, First Out' (LIFO).

Interestingly, even in the case of purely acoustic or elastic sensitivity kernels,
i.e., in the absence of attenuation, Process~C is a little more accurate than Process~B
in terms of numerical errors, because in~C one computes the exact same forward run twice, the second time from
intermediate restart files, thus resetting numerical errors and getting cumulated numerical dispersion for a total of~$N$ time steps only,
while in~B one performs the forward simulation and then a second forward simulation backwards
from the saved snapshot of the final time step, thus getting cumulated numerical dispersion for a total of~$2N$ time steps.

\section{Validation benchmark}

In this section we validate our approach to compute anelastic sensitivity kernels (Process~C shown in Figure~\ref{how_we_can_undo_attenuation}c)
by comparing its results to those obtained with the exact approach (Process~A shown in Figure~\ref{how_we_can_undo_attenuation}a).
In order to illustrate the effects of full attenuation on sensitivity kernels,
we also compare our kernels to an approximate kernel in which only physical dispersion is taken into account,
i.e., Process~B shown in Figure~\ref{how_we_can_undo_attenuation}b, but with a total of four runs performed instead of three,
as discussed in Section~\ref{introduction}. We calculate kernels for a cross-correlation traveltime measurement,
i.e., we use time-reversed particle velocity as the adjoint source, as explained in~\cite{TrTaLi05}.

In order to perform the benchmark, we resort to the spectral-element method~\citep[e.g.,][]{KoTr99,KoTr02a}.
The mesh of hexahedra used in the 3D simulations is designed to honor all first-order discontinuities
in the Preliminary Reference Earth Model (PREM)~\citep{DzAn81},
which are the Moho at a depth of 24.4~km, the upper mantle discontinuities
at depths of 220~km, 400~km, and 670~km, the core-mantle boundary, and the inner-core boundary; it also
honors second-order discontinuities at 600~km, 771~km, and at the top of the D'' layer.
The mesh is doubled in size once below the Moho, a second time below the 670~km discontinuity, and a third time in the middle of the outer core~\citep{KoTr02a}.
Each of the six chunks that comprise the so-called `cubed sphere' that the spectral-element technique uses to mesh the Earth
has 256~$\times$~256 elements along the free surface
and, as a result of the three doublings, 32~$\times$~32 elements along the inner-core boundary,
leading to a total of 4,352,000 spectral elements to mesh the entire globe.
The radial density and velocity profiles of the model are determined by PREM.
The 3-km thick water layer of PREM has been replaced with the PREM upper crust.
PREM has a transversely isotropic asthenosphere between 24.4~km and 220~km,
which is also incorporated in our simulations.
Based on the size of the mesh cells, the simulations presented in this section are accurate for periods greater than about 17~s.
To ensure stability and accuracy of the calculations we use a time step~$\Delta t$ = 0.19~s.
We simulate a total duration of 5,400~s, i.e., 28,600 time steps.
We resort to parallel computing using a total of 384 processor cores.

A source with a source time function with a half duration of 11.2~s, strike of
$174^\circ$, dip of~$30^\circ$, and rake of~$67^\circ$ is located at latitude
$-16.08^\circ$ and longitude~$168.31^\circ$, at a depth of 15~km (corresponding
to event 112699G in the global CMT catalog). A receiver is located on the
surface of the Earth at latitude~$25.10^\circ$ and longitude~$52.37^\circ$, at
an epicentral distance of~$120^\circ$, and records the three components of the displacement vector.

\begin{figure}
\centerline{\includegraphics[width=\linewidth]{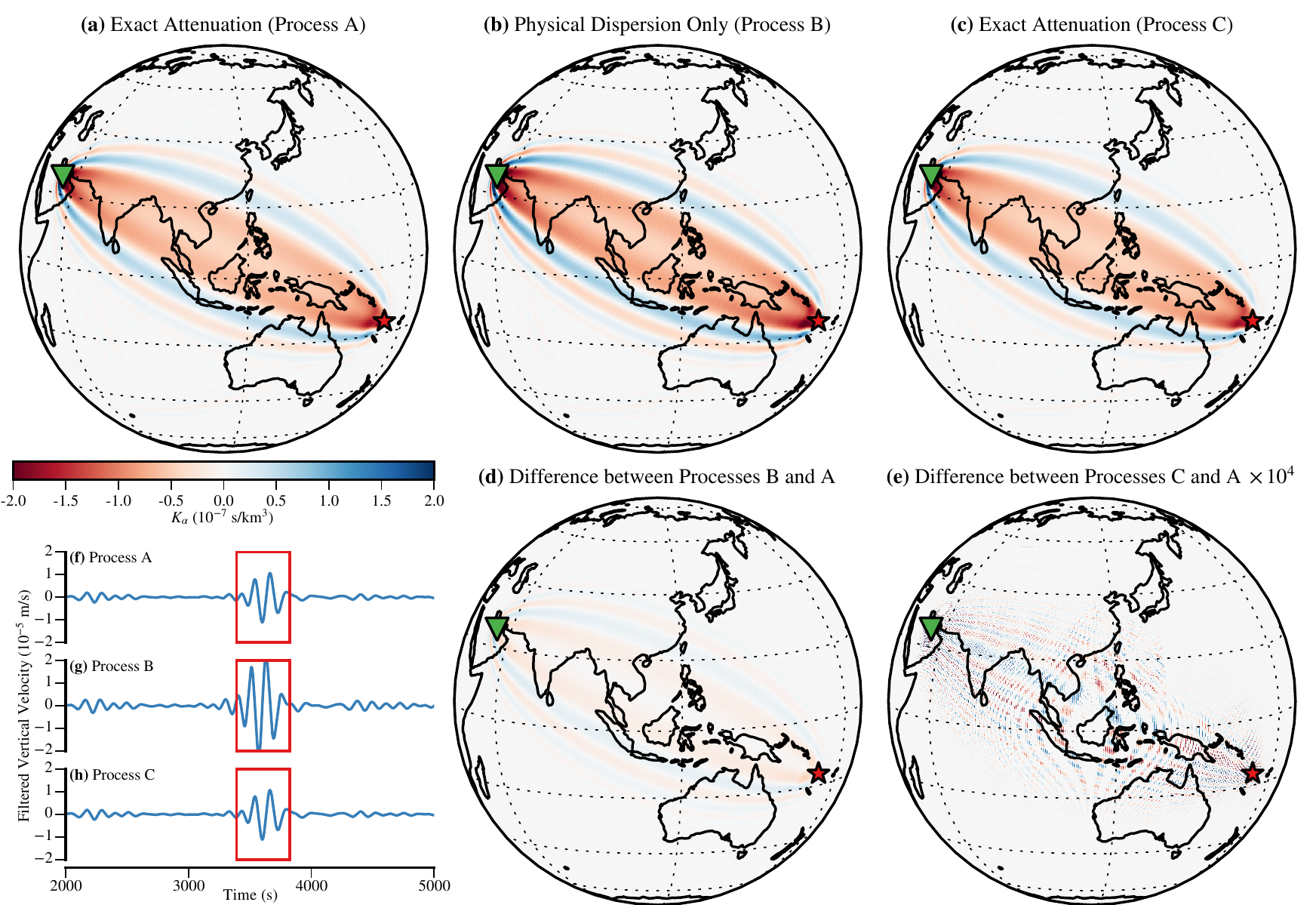}}
\caption{\small %
(a)~Reference exact anelastic~$K_\alpha$ sensitivity kernel defined by
  equation~(\ref{definition_of_Kalpha}) in the 100--200~s period range at
  an epicentral distance of 120$^\circ$ obtained by saving the entire forward
  simulation to disk. (Process~A in Figure~\ref{how_we_can_undo_attenuation}a.) The
  source is indicated by the red star and the receiver by the green triangle.
(b)~Approximate anelastic~$K_\alpha$ sensitivity kernel obtained by
  taking into account physical dispersion only, as explained in
  Section~\ref{introduction}. (Process~A in Figure~\ref{how_we_can_undo_attenuation}a.)
  This kernel matches the exact result in terms of its pattern, but differs on average by~30\% in magnitude.
(c)~Anelastic sensitivity kernel computed with our new method.
  (Process~C in Figure~\ref{how_we_can_undo_attenuation}c.) This kernel
  matches the exact result in both pattern and magnitude, with differences of
  less than~0.01\%.
(d)~The difference between Processes~B and A.
(e)~The difference between Processes~C and A, enhanced by a factor of~$10^4$.
(f-h)~Vertical component synthetic seismograms filtered in the 100--200~s passband.
  The red rectangle indicates the surface wave signal used to create the
  adjoint source. The seismograms obtained by the exact and new methods are
  identical since the forward simulation is the same in both cases. The
  seismogram obtained using physical dispersion only shows the
  surface-wave arriving 30~s early on average, with double the peak amplitude.}
\label{attenuation-comparison}
\end{figure}

Figure~\ref{attenuation-comparison} shows
the~$K_\alpha$ sensitivity kernel defined by equation~(\ref{definition_of_Kalpha}) in the 100--200~s period range
obtained based on Processes~A, B, and C.
The difference between Processes~A and C is shown in Figure~\ref{attenuation-comparison}e and confirms that our technique works well and is exact.
When using physical dispersion in Process~B, the difference in the resulting kernel is on average~30\% of the exact result,
as shown in Figure~\ref{attenuation-comparison}d.
The non-negligible differences that appear when only physical dispersion is taken into account rather than full attenuation
highlight the importance of including full attenuation in the calculation of sensitivity kernels,
as discussed further in the next section.
Note that the convention for traveltime kernels throughout this paper is such that
$\Delta T = T^\mathrm{obs} - T^\mathrm{syn}$, where $T^\mathrm{obs}$ and $T^\mathrm{syn}$ are the travel times of observed and synthetic data
\citep[e.g.,][]{Marq99,DaBa02}.

We have chosen to represent the~$K_\alpha$ kernel because, as shown in equation~(\ref{definition_of_Kalpha}),
it requires saving a single scalar to disk during Stage~1 of Process~A, namely the trace of the strain tensor,
$\bdel \cdot \bs$, as a function of time and space. Saving that scalar required 16~TB of disk space,
thus illustrating that Process~A is currently inconvenient and
cannot be routinely used when conducting seismic imaging at relatively high frequencies.
In comparison, computing the exact~$K_\beta$ kernel given by equation~(\ref{definition_of_Kbeta})
via Process~A would require 50~TB of additional disk storage.

\section{Importance of full attenuation in kernel calculations}
\label{SectionImportance}

In this section we compare sensitivity kernels calculated based on physical dispersion only (Process~B) with exact kernels calculated based on
our new parsimonious storage technique (Process~C).
As discussed in detail by~\cite{ZhLiTr11},
for body-wave traveltime measurements attenuation can be safely ignored, as long as the kernels are calculated in models with the
appropriate wavespeed, that is, taking into account the effects of physical dispersion
defined in equation~(\ref{eq:PD}) and illustrated in Figure~\ref{taking_into_account_average_physical_dispersion}.
\cite{ZhLiTr11} also show that for intermediate-period surface waves at regional distances the physical-dispersion-only approach is perfectly valid.
Thus the use of Process~B to accommodate the effects of attenuation
in regional-scale studies \citep[e.g.,][]{ZhBoPeTr12,ZhTr13,ChenNiuLiuTr2015} is well justified.

Process~B may be summarized as follows:
\begin{enumerate}
\item Compute two sets of synthetic seismograms, 1) using full attenuation, and 2) using physical dispersion only.
\item Make measurements between observed and synthetic seismograms with full attenuation.
\item Calculate sensitivity kernels for this measurement using physical-dispersion-only forward and adjoint wavefields.
\item Compute the gradient by weighting the kernels obtained in step~(iii) with the measurements from step~(ii).
\end{enumerate}
Note that process~B is suitable for cross-correlation or frequency-dependent (e.g., multi-taper) traveltime and amplitude anomaly measurements,
but nor for FWI. In contrast, the new approach based on Process~C, taking into account full attenuation, may be summarized as follows:
\begin{enumerate}
\item Compute synthetic seismograms using full attenuation.
\item Make measurements between observed and synthetic seismograms with full attenuation.
\item Calculate sensitivity kernels for this measurement using forward and adjoint wavefields with full attenuation based on Process~C.
\item Compute the gradient by weighting the kernels obtained in step~(iii) with the measurements from step~(ii).
\end{enumerate}
The number of simulations required for Process~C is reduced by a factor~4/3 compared to Process~B,
although the extra cost is partially offset by relatively cheaper physical dispersion-only simulations in Process~B.

\begin{figure}
  \center
  \includegraphics[width=\textwidth]{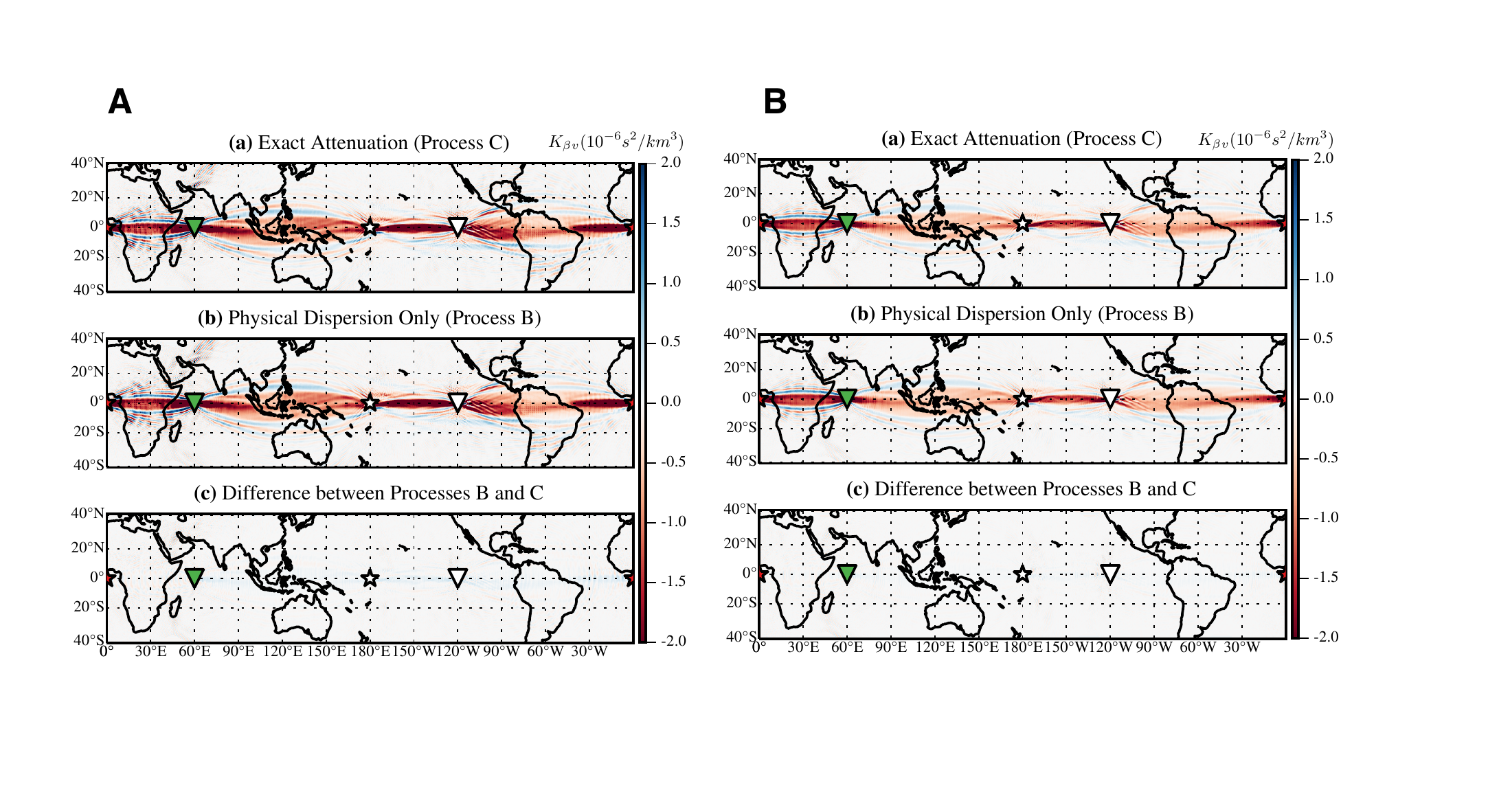}
  \caption{\small Multi-taper traveltime shear wavespeed sensitivity kernels~$K_{\beta_{v}}$
for 40--60~s vertical-component  R1 \& R2 waves at depths of A) 30 km and B) 125 km.
The minor-arc epicentral distance is~$60^\circ$.
Traveltime measurements are set to -1 in the computations (i.e., $\Delta T(\omega)=-1$).
The locations of the source and receiver are indicated by the red star and green triangle, respectively.
The white star and triangle denote the source and receiver antipodes, respectively.
S40RTS with Crust2.0 is used as the 3D model to compute forward and adjoint simulations.
Associated seismograms and adjoint sources are as shown in Figure~\ref{fig:waveforms_Rayleigh}.}
  \label{fig:Rayleigh}
\end{figure}

\begin{figure}
  \center
\vspace*{-2truecm}
  \includegraphics[width=0.88\textwidth]{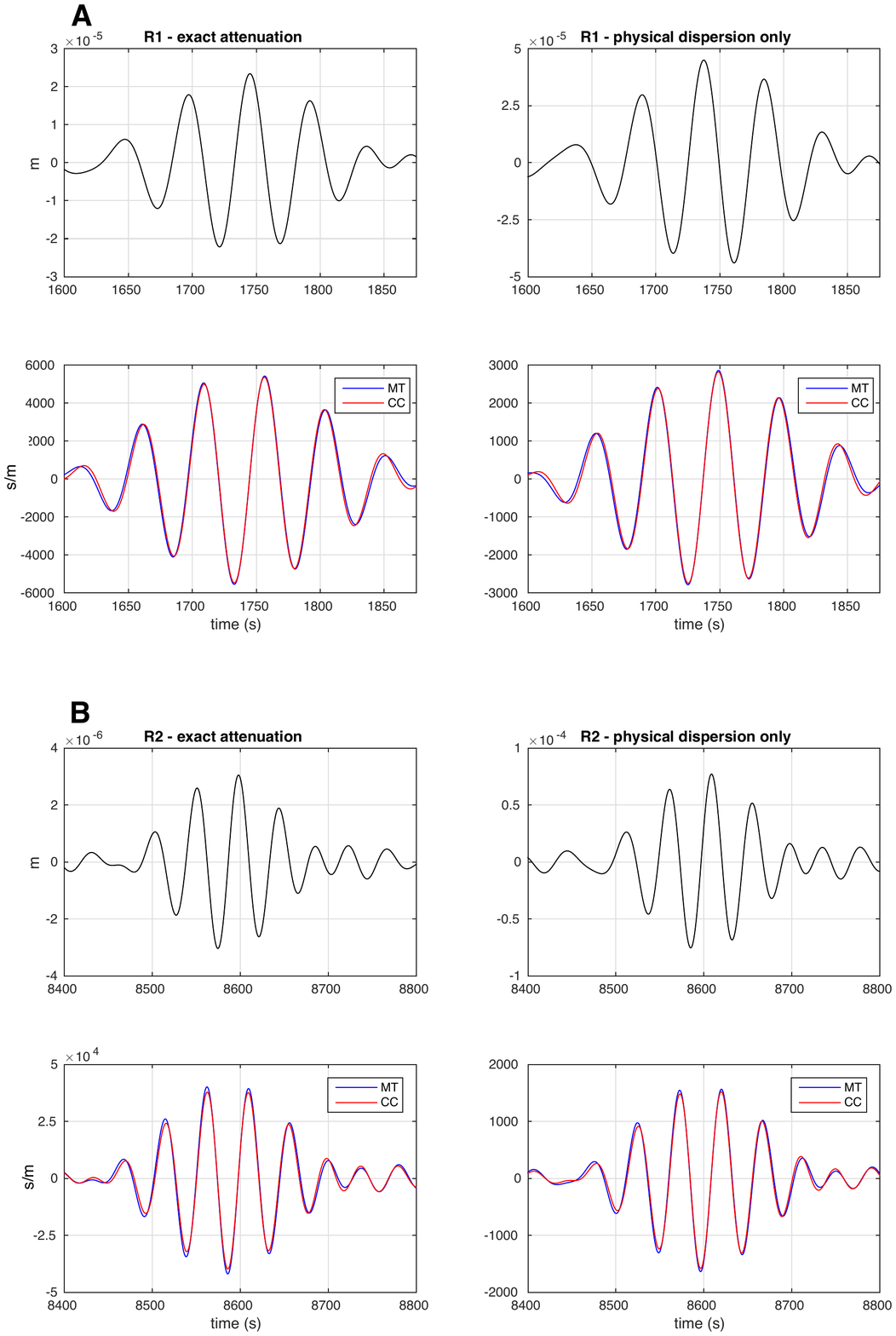}
\vspace*{-1.truecm}
 \caption{\small A) Vertical-component R1 seismograms computed with physical dispersion only and full attenuation ({\it top row})
and their associated adjoint sources ({\it bottom row}).
B) Vertical-component R2 seismograms computed with physical dispersion only and full attenuation ({\it top row}) and their associated adjoint sources,
where the measurements (i.e., $\Delta T $) are set to -1 ({\it bottom row}).
CC and MT denote cross-correlation traveltime and multi-taper measurements, respectively.
The multi-taper adjoint sources are used to compute
the~$K_{\beta_{v}}$ kernels presented in Figure~\ref{fig:Rayleigh}. The epicentral distance is~$60^\circ$, and seismograms were filtered between 40--60~s.
S40RTS with Crust2.0 is used as the 3D model to compute the seismograms. Notice that, due to the relatively narrow-band signals, the physical-dispersion-only
and full-attenuation waveforms are similar.
Note also that, due to the nondispersive behavior of the wavetrains, the CC and MT adjoint sources are very similar.}
  \label{fig:waveforms_Rayleigh}
\end{figure}

In Figure~\ref{fig:Rayleigh}, we present horizontal cross-sections of multi-taper traveltime shear-wavespeed sensitivity kernels
defined by equation~(\ref{definition_of_Kbeta}) at depths of 30~km and 125~km
for 40--60~s vertical-component Rayleigh waves using physical-dispersion-only and full attenuation, respectively. We used 3D mantle model S40RTS
\cite{RiDeVaWo10} together with 3D crustal model Crust2.0 \cite{BaLaMa00} as a background model during forward and adjoint simulations.
Confirming observations by~\cite{ZhLiTr11}, the two sets of kernels are in good agreement.
The corresponding R1 and R2 seismograms together with their adjoint sources computed based on cross-correlation and multi-taper measurements are shown in
Figure~\ref{fig:waveforms_Rayleigh}.
The difference between physical-dispersion-only and full-attenuation kernels is mainly in amplitude,
although the former exhibit slight differences in shape compared to the latter.
The success of physical-dispersion-only kernels strongly depends on the choice of measurement and the bandpass.
As long as the physical-dispersion-only and full attenuation waveforms are similar in shape, the resulting kernels will also be similar, as is clearly shown
for 40~s Rayleigh waves.

\begin{figure}
  \center
  \includegraphics[width=\textwidth]{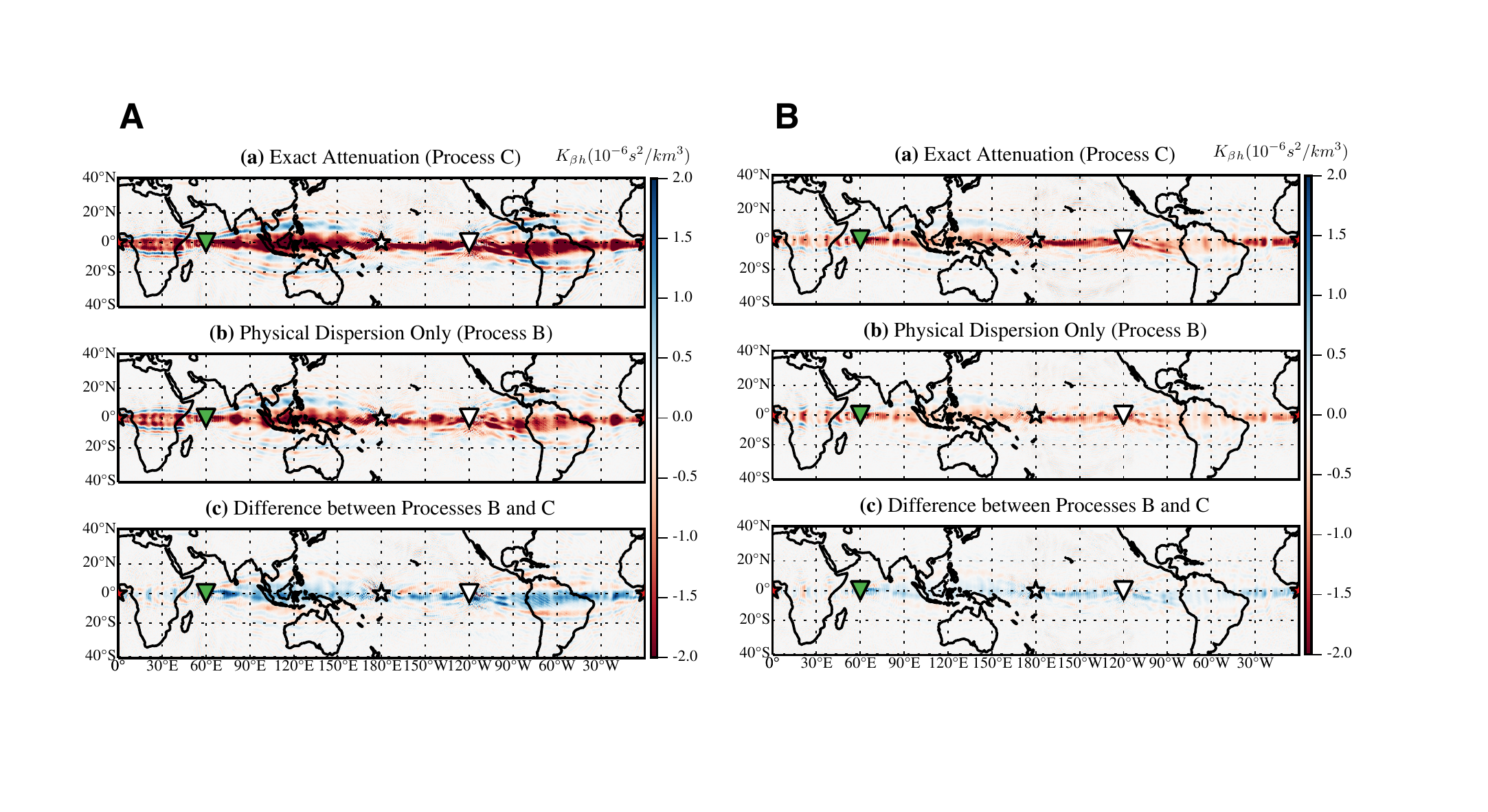}
  \caption{\small Multi-taper traveltime shear wavespeed sensitivity kernels~$K_{\beta_{h}}$ for 40--60~s transverse-component  G1 \& G2 waves at depths of A) 30 km
and B) 125 km. The minor-arc epicentral distance is~$60^\circ$.
Traveltime measurements are set to -1 in the computations (i.e., $\Delta T(\omega)=-1$).
The locations of the source and receiver are shown by the red star and the green triangle,
respectively. The white star and triangle denote the source and receiver antipodes, respectively. S40RTS with Crust2.0 is used as the 3D model to compute
forward and adjoint simulations. Associated seismograms and adjoint sources are as shown in Figure~\ref{fig:waveforms_Love}.}
  \label{fig:Love}
\end{figure}

\begin{figure}
  \center
\vspace*{-2truecm}
  \includegraphics[width=0.88\textwidth]{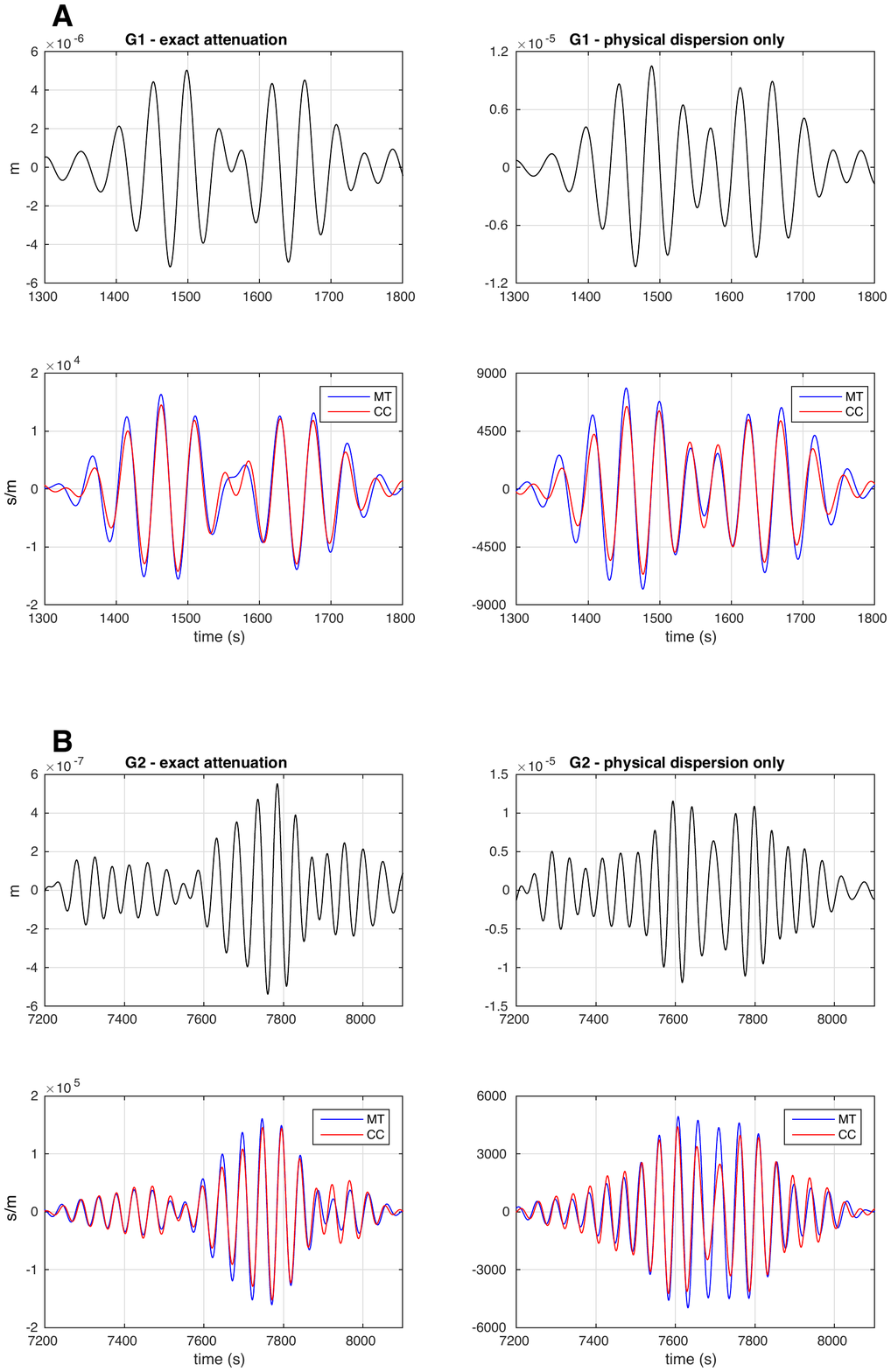}
\vspace*{-1.truecm}
 \caption{\small A) Transverse-component G1 seismograms computed with physical dispersion only and full attenuation ({\it top row})
and their associated adjoint sources ({\it bottom row}).
B) Transverse-component G2 seismograms computed with physical dispersion only and full attenuation ({\it top row}) and their associated adjoint sources,
where the measurements (i.e., $\Delta T $) are set to -1 ({\it bottom row}).
CC and MT denote cross-correlation traveltime and multi-taper measurements, respectively.
The multi-taper adjoint sources are used to compute
the~$K_{\beta{h}}$ kernels presented in Figure~\ref{fig:Love}. The epicentral distance is~$60^\circ$, and seismograms were filtered between 40--60~s. S40RTS
with Crust2.0 is used as the 3D model to compute the seismograms. Notice that we start observing the effect of full attenuation more for Love waves due to
their higher sensitivity to crustal variations.}
  \label{fig:waveforms_Love}
\end{figure}

In Figure~\ref{fig:Love}, we present horizontal cross-sections of Love-wave multi-taper traveltime shear wavespeed sensitivity kernels
defined by equation~(\ref{definition_of_Kbeta}) at 30~km and 125~km depths.
Shown are 40--60~s transverse-component Love-wave kernels using physical-dispersion-only and full attenuation, respectively. Again, S40RTS \cite{RiDeVaWo10}
together with Crust2.0 \cite{BaLaMa00} is used during the numerical simulations. The corresponding G1 and G2 seismograms together with their adjoint sources
computed based on cross-correlation and multi-taper measurements are shown in Figure~\ref{fig:waveforms_Love}. For these shorter period Love waves we see that
the physical dispersion-only sensitivity kernels are beginning to break down,
especially along the major arc, mainly due to their stronger sensitivity to the 3D crustal heterogeneity.

\section{Conclusions and future work}

We have introduced a method of computing exact anelastic sensitivity kernels in the time domain
using parsimonious disk storage and a simple reordering of the time loop,
combined with the use of a `Last In, First Out' memory buffer.
The total number of time steps required is unaffected compared to usual approaches for the acoustic or elastic (nondissipative) cases.
We reduced the computational cost by a factor~4/3 compared to a commonly used approach in which only the effects of physical dispersion associated with
anelasticity are taken into account.

We performed a benchmark in which we compared the compressional wavespeed sensitivity kernel
obtained based on our approach to the exact
kernel obtained by saving the entire forward calculation to disk; the difference was zero, confirming that our approach is also exact.
For shorter-period surface waves we discovered non-negligible kernel differences,
thus illustrating the importance of including full attenuation in sensitivity kernel calculations for dispersive waves.

The technique applies without modification to problems in reverse-time migration,
which may be viewed as a particular case of a sensitivity kernel calculation
\citep[e.g.,][]{ViOp09,DoYiVaTr10}, time-reversal seismological source studies~\citep[e.g.,][]{LaMoFiCaToCl06},
and time reversal as used in medical imaging or non-destructive testing~\citep[e.g.,][]{FiCaDePrRoTaThWu00,TaFi14}.
It would work for Maxwell's equations as well, since they can be written as a hyperbolic system and are also self-adjoint in the absence of dissipation.
The technique works particularly well on GPU-accelerated machines~\citep[e.g.,][]{KoErGoMi10,Kom11},
because the entire memory of the CPU is largely unused and thus available as a huge memory buffer.

Our SPECFEM open source spectral-element software package is freely available
via the Computational Infrastructure for Geodynamics (CIG; \texttt{geodynamics.org}), including the new developments presented in this article.

\section*{Acknowledgments}
We thank Mark Asch, Didier Auroux, C\'edric Bellis, \'Elie Bretin, Andreas Fichtner, Josselin Garnier, Thomas Guillet, Ioannis G.\ Kevrekidis,
Bruno Lombard, Vadim Monteiller, and William W. Symes for fruitful discussion,
and the Computational Infrastructure for Geodynamics (CIG) and Marie Cournille for support.
We thank Heiner Igel and an anonymous reviewer for useful comments that improved the manuscript.
Part of this work was funded by the Simone and Cino del Duca/Institut de France/French Academy of Sciences Foundation under grant \#095164,
by the European Union Horizon 2020 Marie Curie Action \#641943 project `WAVES' of call H2020-MSCA-ITN-2014,
by U.S.\ NSF grant 1112906 and by China NSFC grant~51378479.
Z.X.\ thanks the China Scholarship Council for financial support during his stay at LMA CNRS, and the continuous support from Prof.\ Liao Zhenpeng.
E.S.\ and Q.L.\ were supported by the NSERC G8 Research Councils Initiative on Multilateral Research Grant No.~490919 and Discovery Grant No.~487237.
This work was granted access to the European Partnership for Advanced Computing in Europe (PRACE) under allocation TGCC CURIE \#ra2410,
to the French HPC resources of TGCC under allocation \#2015-gen7165
made by GENCI and of the Aix-Marseille Supercomputing Mesocenter under allocations \#14b013 and \#15b034,
to the Oak Ridge Leadership Computing Facility at Oak Ridge National Laboratory, USA,
which is supported by the Office of Science of the U.S. Department of Energy under Contract No.~DE-AC05-00OR22725,
and to the Sandybridge cluster at the SciNet HPC Consortium funded by the Canada
Foundation for Innovation, the Ontario Research Fund, and the University of Toronto Startup Fund.
Part of this work was presented at the GPU'2014 Conference in Roma, Italy, in September 2014.

%%%%
%%%%  Bibliography
%%%%

\begin{thebibliography}{70}
\expandafter\ifx\csname natexlab\endcsname\relax\def\natexlab#1{#1}\fi

\bibitem[Ak{\c c}elik et~al.(2003)Ak{\c c}elik, Bielak, Biros, Epanomeritakis,
  Fernandez, Ghattas, Kim, Lopez, O'Hallaron, Tu, \&
  Urbanic]{AkBiBiEpFeGhKiLoOhTuUr03}
Ak{\c c}elik, V., Bielak, J., Biros, G., Epanomeritakis, I., Fernandez, A.,
  Ghattas, O., Kim, E.~J., Lopez, J., O'Hallaron, D., Tu, T., \& Urbanic, J.,
  2003.
\newblock High-resolution forward and inverse earthquake modeling on terascale
  computers, in {\em Proceedings of the SC'03 ACM/IEEE conference on
  Supercomputing\/}, pp. 52--72, Phoenix, Arizona, USA.

\bibitem[Ammari et~al.(2013)Ammari, Bretin, Garnier, \& Wahab]{AmBrGaWa13}
Ammari, H., Bretin, E., Garnier, J., \& Wahab, A., 2013.
\newblock Time-reversal algorithms in viscoelastic media, {\it European Journal
  of Applied Mathematics\/}, {\bf 24}(4), 565--600.

\bibitem[Anderson et~al.(2012)Anderson, Tan, \& Wang]{AnTaWa12}
Anderson, J.~E., Tan, L., \& Wang, D., 2012.
\newblock Time-reversal checkpointing methods for {RTM} and {FWI}, {\it
  Geophysics\/}, {\bf 77}(4), S93--S103.

\bibitem[Askan et~al.(2007)Askan, Ak{\c c}elik, Bielak, \& Ghattas]{AsAkBiGh07}
Askan, A., Ak{\c c}elik, V., Bielak, J., \& Ghattas, O., 2007.
\newblock Full waveform inversion for seismic velocity and anelastic losses in
  heterogeneous structures, {\it Bull. seism. Soc. Am.\/}, {\bf 97}(6),
  1990--2008.

\bibitem[Assimaki et~al.(2012)Assimaki, Kallivokas, Kang, Li, \&
  Kucukcoban]{AsKaKaLiKu12}
Assimaki, D., Kallivokas, L.~F., Kang, J.~W., Li, W., \& Kucukcoban, S., 2012.
\newblock Time-domain forward and inverse modeling of lossy soils with
  frequency-independent {$Q$} for near-surface applications, {\it Soil Dynamics
  and Earthquake Engineering\/}, {\bf 43}, 139--159.

\bibitem[Auroux et~al.(2011)Auroux, Blum, \& Nodet]{AuBlNo11}
Auroux, D., Blum, J., \& Nodet, M., 2011.
\newblock Diffusive back and forth nudging algorithm for data assimilation,
  {\it C. R. Acad. Sci. Paris, Ser. I\/}, {\bf 349}(15-16), 849--854.

\bibitem[Auroux et~al.(2013)Auroux, Bansart, \& Blum]{AuBaBl13}
Auroux, D., Bansart, P., \& Blum, J., 2013.
\newblock An evolution of the back and forth nudging for geophysical data
  assimilation: application to {B}urgers equation and comparisons, {\it Inverse
  Problems in Science and Engineering\/}, {\bf 21}(3), 399--419.

\bibitem[Bassin et~al.(2000)Bassin, Laske, \& Masters]{BaLaMa00}
Bassin, C., Laske, G., \& Masters, G., 2000.
\newblock The current limits of resolution for surface wave tomography in
  {N}orth {A}merica, {\it EOS\/}, {\bf 81}, F897.

\bibitem[Blanc et~al.(2016)Blanc, Komatitsch, Chaljub, Lombard, \&
  Xie]{BlKoChLoXi16}
Blanc, E., Komatitsch, D., Chaljub, E., Lombard, B., \& Xie, Z., 2016.
\newblock Highly accurate stability-preserving optimization of the {Z}ener
  viscoelastic model, with application to wave propagation in the presence of
  strong attenuation, {\it Geophys. J. Int.\/}, {\bf 205}(1), 427--439.

\bibitem[Carcione(2014)]{Car14}
Carcione, J.~M., 2014.
\newblock {\it Wave fields in real media: {W}ave propagation in anisotropic,
  anelastic, porous and electromagnetic media\/}, Elsevier Science, Amsterdam,
  The Netherlands, 3rd edn.

\bibitem[Charpentier(2001)]{Cha01}
Charpentier, I., 2001.
\newblock Checkpointing schemes for adjoint codes: {A}pplication to the
  meteorological model {Meso-NH}, {\it SIAM Journal on Scientific Computing\/},
  {\bf 22}(6), 2135--2151.

\bibitem[Chen et~al.(2015)Chen, Niu, Liu, Tromp, \& Zheng]{ChenNiuLiuTr2015}
Chen, M., Niu, F., Liu, Q., Tromp, J., \& Zheng, X., 2015.
\newblock Multi parameter adjoint tomography of the crust and upper mantle
  beneath {E}ast {A}sia: 1. {M}odel construction and comparisons, {\it J.
  Geophys. Res.\/}, {\bf 120}, 1762--1786.

\bibitem[Cyr et~al.(2015)Cyr, Shadid, \& Wildey]{CyShWi15}
Cyr, E.~C., Shadid, J.~N., \& Wildey, T., 2015.
\newblock Towards efficient backward-in-time adjoint computations using data
  compression techniques, {\it Comput. Methods Appl. Mech. Eng.\/}, {\bf 288},
  24--44.

\bibitem[Dahlen \& Baig(2002)]{DaBa02}
Dahlen, F.~A. \& Baig, A.~M., 2002.
\newblock Fr\'echet kernels for body-wave amplitudes, {\it Geophys. J. Int.\/},
  {\bf 150}, 440--466.

\bibitem[Dahlen \& Tromp(1998)]{DaTr98}
Dahlen, F.~A. \& Tromp, J., 1998.
\newblock {\it Theoretical Global Seismology\/}, Princeton University Press,
  Princeton, New-Jersey, USA, 944 pages.

\bibitem[Dahlen et~al.(2000)Dahlen, Hung, \& Nolet]{DaHuNo00}
Dahlen, F.~A., Hung, S.-H., \& Nolet, G., 2000.
\newblock Fr{\'e}chet kernels for finite-frequency traveltimes - {I. T}heory,
  {\it Geophys. J. Int.\/}, {\bf 141}(1), 157--174.

\bibitem[Douma et~al.(2010)Douma, Yingst, Vasconcelos, \& Tromp]{DoYiVaTr10}
Douma, H., Yingst, D., Vasconcelos, I., \& Tromp, J., 2010.
\newblock On the connection between artifact filtering in reverse-time
  migration and adjoint tomography, {\it Geophysics\/}, {\bf 75}(6),
  S219--S223.

\bibitem[Dziewo\'nski \& Anderson(1981)]{DzAn81}
Dziewo\'nski, A.~M. \& Anderson, D.~L., 1981.
\newblock Preliminary reference {E}arth model, {\it Phys. Earth planet.
  Inter.\/}, {\bf 25}(4), 297--356.

\bibitem[Felgenhauer(1992)]{Fel92}
Felgenhauer, U., 1992.
\newblock Quasi-{N}ewton descent methods with inexact gradients, in {\em
  Operations Research '91\/}, pp. 83--86, eds Gritzmann, P., Hettich, R.,
  Horst, R., \& Sachs, E., Physica-Verlag, Heidelberg, Germany.

\bibitem[Fichtner(2010)]{Fic10}
Fichtner, A., 2010.
\newblock {\it Full Seismic Waveform Modelling and Inversion\/}, Advances in
  Geophysical and Environmental Mechanics and Mathematics, Springer-Verlag,
  Berlin, Germany, 343 pages.

\bibitem[Fichtner \& van Driel(2014)]{FiVa14}
Fichtner, A. \& van Driel, M., 2014.
\newblock Models and {F}r\'echet kernels for frequency-(in)dependent {$Q$},
  {\it Geophys. J. Int.\/}, {\bf 198}(3), 1878--1889.

\bibitem[Fichtner et~al.(2006)Fichtner, Bunge, \& Igel]{FiBuIg06}
Fichtner, A., Bunge, H.-P., \& Igel, H., 2006.
\newblock The adjoint method in seismology: {I. T}heory, {\it Phys. Earth
  planet. Inter.\/}, {\bf 157}(1-2), 86--104.

\bibitem[Fichtner et~al.(2009)Fichtner, Kennett, Igel, \& Bunge]{FiKeIg09}
Fichtner, A., Kennett, B. L.~N., Igel, H., \& Bunge, H.~P., 2009.
\newblock Full seismic waveform tomography for upper-mantle structure in the
  {A}ustralasian region using adjoint methods, {\it Geophys. J. Int.\/}, {\bf
  179}(3), 1703--1725.

\bibitem[Fink et~al.(2000)Fink, Cassereau, Derode, Prada, Roux, Tanter, Thomas,
  \& Wu]{FiCaDePrRoTaThWu00}
Fink, M., Cassereau, D., Derode, A., Prada, C., Roux, P., Tanter, M., Thomas,
  J.-L., \& Wu, F., 2000.
\newblock Time-reversed acoutics, {\it Reports on Progress in Physics\/}, {\bf
  63}(12), 1933--1995.

\bibitem[Griewank \& Walther(2000)]{GrWa00}
Griewank, A. \& Walther, A., 2000.
\newblock Algorithm 799: {R}evolve, an implementation of checkpointing for the
  reverse or adjoint mode of computational differentiation, {\it ACM
  Transactions on Mathematical Software\/}, {\bf 26}(1), 19--45.

\bibitem[Groos et~al.(2014)Groos, Sch\"afer, Forbriger, \& Bohlen]{GrScFoBo14}
Groos, L., Sch\"afer, M., Forbriger, T., \& Bohlen, T., 2014.
\newblock The role of attenuation in {2D} full-waveform inversion of
  shallow-seismic body and {R}ayleigh waves, {\it Geophysics\/}, {\bf 79}(6),
  R247--R261.

\bibitem[Hinze et~al.(2006)Hinze, Walther, \& Sternberg]{HiWaSt06}
Hinze, M., Walther, A., \& Sternberg, J., 2006.
\newblock An optimal memory-reduced procedure for calculating adjoints of the
  instationary {N}avier-{S}tokes equations, {\it Optimal Control Applications
  and Methods\/}, {\bf 27}(1), 19--40.

\bibitem[Kallivokas et~al.(2013)Kallivokas, Fathi, Kucukcoban, {Stokoe II},
  Bielak, \& Ghattas]{KaFaKuStBiGh13}
Kallivokas, L.~F., Fathi, A., Kucukcoban, S., {Stokoe II}, K.~H., Bielak, J.,
  \& Ghattas, O., 2013.
\newblock Site characterization using full waveform inversion, {\it Soil
  Dynamics and Earthquake Engineering\/}, {\bf 47}, 62--82.

\bibitem[Komatitsch(2011)]{Kom11}
Komatitsch, D., 2011.
\newblock Fluid-solid coupling on a cluster of {GPU} graphics cards for seismic
  wave propagation, {\it C. R. Acad. Sci., Ser. IIb Mec.\/}, {\bf 339},
  125--135.

\bibitem[Komatitsch \& Tromp(1999)]{KoTr99}
Komatitsch, D. \& Tromp, J., 1999.
\newblock Introduction to the spectral-element method for 3-{D} seismic wave
  propagation, {\it Geophys. J. Int.\/}, {\bf 139}(3), 806--822.

\bibitem[Komatitsch \& Tromp(2002)]{KoTr02a}
Komatitsch, D. \& Tromp, J., 2002.
\newblock Spectral-element simulations of global seismic wave propagation{-I.
  V}alidation, {\it Geophys. J. Int.\/}, {\bf 149}(2), 390--412.

\bibitem[Komatitsch \& Vilotte(1998)]{KoVi98}
Komatitsch, D. \& Vilotte, J.~P., 1998.
\newblock The spectral-element method: an efficient tool to simulate the
  seismic response of 2{D} and 3{D} geological structures, {\it Bull. seism.
  Soc. Am.\/}, {\bf 88}(2), 368--392.

\bibitem[Komatitsch et~al.(2010)Komatitsch, Erlebacher, G\"oddeke, \&
  Mich\'ea]{KoErGoMi10}
Komatitsch, D., Erlebacher, G., G\"oddeke, D., \& Mich\'ea, D., 2010.
\newblock High-order finite-element seismic wave propagation modeling with
  {MPI} on a large {GPU} cluster, {\it J. Comput. Phys.\/}, {\bf 229}(20),
  7692--7714.

\bibitem[Kowar \& Scherzer(2011)]{KoSc11}
Kowar, R. \& Scherzer, O., 2011.
\newblock Photoacoustic imaging taking into account attenuation, in {\em
  Mathematical Modeling in Biomedical Imaging II, Lecture Notes in
  Mathematics\/}, vol. 2035, pp. 85--130, Springer-Verlag, Berlin, Germany.

\bibitem[Kurzmann et~al.(2013)Kurzmann, Przebindowska, K\"ohn, \&
  Bohlen]{KuPrKoBo13}
Kurzmann, A., Przebindowska, A., K\"ohn, D., \& Bohlen, T., 2013.
\newblock Acoustic full waveform tomography in the presence of attenuation: a
  sensitivity analysis, {\it Geophys. J. Int.\/}, {\bf 195}(2), 985--1000.

\bibitem[Larmat et~al.(2006)Larmat, Montagner, Fink, Capdeville, Tourin, \&
  Cl\'ev\'ed\'e]{LaMoFiCaToCl06}
Larmat, C., Montagner, J.-P., Fink, M., Capdeville, Y., Tourin, A., \&
  Cl\'ev\'ed\'e, E., 2006.
\newblock Time-reversal imaging of seismic sources and application to the great
  {S}umatra earthquake, {\it Geophys. Res. Lett.\/}, {\bf 33}(19), L19312.

\bibitem[Liu et~al.(1976)Liu, Anderson, \& Kanamori]{LiAnKa76}
Liu, H.~P., Anderson, D.~L., \& Kanamori, H., 1976.
\newblock Velocity dispersion due to anelasticity: implications for seismology
  and mantle composition, {\it Geophys. J. Roy. Astron. Soc.\/}, {\bf 47},
  41--58.

\bibitem[Liu \& Tromp(2006)]{LiTr06}
Liu, Q. \& Tromp, J., 2006.
\newblock Finite-frequency kernels based on adjoint methods, {\it Bull. seism.
  Soc. Am.\/}, {\bf 96}(6), 2383--2397.

\bibitem[Liu \& Tromp(2008)]{LiTr08}
Liu, Q. \& Tromp, J., 2008.
\newblock Finite-frequency sensitivity kernels for global seismic wave
  propagation based upon adjoint methods, {\it Geophys. J. Int.\/}, {\bf
  174}(1), 265--286.

\bibitem[Marquering et~al.(1999)Marquering, Dahlen, \& Nolet]{Marq99}
Marquering, H., Dahlen, F.~A., \& Nolet, G., 1999.
\newblock Three-dimensional sensitivity kernels for finite-frequency
  traveltimes: the banana-doughnut paradox, {\it Geophys. J. Int.\/}, {\bf
  137}(3), 805--815.

\bibitem[Monteiller et~al.(2015)Monteiller, Chevrot, Komatitsch, \&
  Wang]{MoChKoWa15}
Monteiller, V., Chevrot, S., Komatitsch, D., \& Wang, Y., 2015.
\newblock Three-dimensional full waveform inversion of short-period teleseismic
  wavefields based upon the {SEM-DSM} hybrid method, {\it Geophys. J. Int.\/},
  {\bf 202}(2), 811--827.

\bibitem[Pakravan et~al.(2016)Pakravan, Kang, \& Newtson]{PaKaNe16}
Pakravan, A., Kang, J.~W., \& Newtson, C.~M., 2016.
\newblock A {G}auss-{N}ewton full-waveform inversion for material profile
  reconstruction in viscoelastic semi-infinite solid media, {\it Inverse
  Problems in Science and Engineering\/}, {\bf 24}(3), 393--421.

\bibitem[Peter et~al.(2011)Peter, Komatitsch, Luo, Martin, {Le Goff},
  Casarotti, {Le Loher}, Magnoni, Liu, Blitz, Nissen-Meyer, Basini, \&
  Tromp]{PeKoLuMaLeCaLeMaLiBlNiBaTr11}
Peter, D., Komatitsch, D., Luo, Y., Martin, R., {Le Goff}, N., Casarotti, E.,
  {Le Loher}, P., Magnoni, F., Liu, Q., Blitz, C., Nissen-Meyer, T., Basini,
  P., \& Tromp, J., 2011.
\newblock Forward and adjoint simulations of seismic wave propagation on fully
  unstructured hexahedral meshes, {\it Geophys. J. Int.\/}, {\bf 186}(2),
  721--739.

\bibitem[Plessix(2006)]{Plessix_2006_RAS}
Plessix, R.~E., 2006.
\newblock A review of the adjoint-state method for computing the gradient of a
  functional with geophysical applications, {\it Geophys. J. Int.\/}, {\bf
  167}(2), 495--503.

\bibitem[Restrepo et~al.(1998)Restrepo, Leaf, \& Griewank]{ReLeGr98}
Restrepo, J.~M., Leaf, G.~K., \& Griewank, A., 1998.
\newblock Circumventing storage limitations in variational data assimilation
  studies, {\it SIAM J. Sci. Comput.\/}, {\bf 19}(5), 1586--1605.

\bibitem[Ritsema et~al.(2011)Ritsema, Deuss, {Van Heijst}, \&
  Woodhouse]{RiDeVaWo10}
Ritsema, J., Deuss, A., {Van Heijst}, H.~J., \& Woodhouse, J.~H., 2011.
\newblock {S40RTS}: a degree-40 shear-velocity model for the mantle from new
  rayleigh wave dispersion, teleseismic traveltime and normal-mode splitting
  function measurements, {\it Geophys. J. Int.\/}, {\bf 184}(3), 1223--1236.

\bibitem[Ruan \& Zhou(2010)]{RuZh10}
Ruan, Y. \& Zhou, Y., 2010.
\newblock The effects of 3-{D} anelasticity ({$Q$}) structure on surface wave
  phase delays, {\it Geophys. J. Int.\/}, {\bf 181}(1), 479--492.

\bibitem[Ruan \& Zhou(2012)]{RuZh12}
Ruan, Y. \& Zhou, Y., 2012.
\newblock The effects of 3-{D} anelasticity ({$Q$}) structure on surface wave
  amplitudes, {\it Geophys. J. Int.\/}, {\bf 189}(2), 967--983.

\bibitem[Rubio~Dalmau et~al.(2014)Rubio~Dalmau, Hanzich, de~la Puente, \&
  Guti\'errez]{RuHaPuGu14}
Rubio~Dalmau, F., Hanzich, M., de~la Puente, J., \& Guti\'errez, N., 2014.
\newblock Lossy data compression with {DCT} transforms, in {\em Proceedings of
  the EAGE Workshop on High Performance Computing for Upstream\/}, p. HPC30,
  Chania, Crete, Greece.

\bibitem[Sherali \& Ulular(1990)]{ShUl90}
Sherali, H.~D. \& Ulular, O., 1990.
\newblock Conjugate gradient methods using quasi-{N}ewton updates with inexact
  line searches, {\it Journal of Mathematical Analysis and Applications\/},
  {\bf 150}(2), 359--377.

\bibitem[Shi(2006)]{Shi06}
Shi, Z.-J., 2006.
\newblock Convergence of quasi-{N}ewton method with new inexact line search,
  {\it Journal of Mathematical Analysis and Applications\/}, {\bf 315}(1),
  120--131.

\bibitem[Spears et~al.(2014)Spears, Corrigan, \& Kailasanath]{SpCoKa14}
Spears, Z., Corrigan, A.~T., \& Kailasanath, K., 2014.
\newblock Checkpointing methods for adjoint-based supersonic jet noise
  reduction, in {\em Proceedings of the 20th AIAA/CEAS Aeroacoustics
  Conference\/}, vol.~1, pp. 694--699, American Institute of Aeronautics and
  Astronautics, Atlanta, Georgia, USA, AIAA paper 2014-2472.

\bibitem[Sun \& Fu(2013)]{SuFu13}
Sun, W. \& Fu, L.-Y., 2013.
\newblock Two effective approaches to reduce data storage in reverse time
  migration, {\it Computers \& Geosciences\/}, {\bf 56}, 69--75.

\bibitem[Symes(2007)]{Sym07}
Symes, W.~W., 2007.
\newblock Reverse time migration with optimal checkpointing, {\it
  Geophysics\/}, {\bf 72}(5), SM213--SM221.

\bibitem[Tanter \& Fink(2014)]{TaFi14}
Tanter, M. \& Fink, M., 2014.
\newblock Ultrafast imaging in biomedical ultrasound, {\it IEEE Transactions on
  Ultrasonics, Ferroelectrics, and Frequency Control\/}, {\bf 61}(1), 102--119.

\bibitem[Tarantola(1986)]{Tarantola_1986_SNL}
Tarantola, A., 1986.
\newblock A strategy for non linear inversion of seismic reflection data, {\it
  Geophysics\/}, {\bf 51}(10), 1893--1903.

\bibitem[Tarantola(1987)]{Tar87}
Tarantola, A., 1987.
\newblock {\it Inverse problem theory: methods for data fitting and model
  parameter estimation\/}, Elsevier Science Publishers, Amsterdam, Netherlands.

\bibitem[Tarantola(1988)]{Tar88}
Tarantola, A., 1988.
\newblock Theoretical background for the inversion of seismic waveforms,
  including elasticity and attenuation, {\it Pure Appl. Geophys.\/}, {\bf 128},
  365--399.

\bibitem[Tromp et~al.(2005)Tromp, Tape, \& Liu]{TrTaLi05}
Tromp, J., Tape, C., \& Liu, Q., 2005.
\newblock Seismic tomography, adjoint methods, time reversal and
  banana-doughnut kernels, {\it Geophys. J. Int.\/}, {\bf 160}(1), 195--216.

\bibitem[Tromp et~al.(2008)Tromp, Komatitsch, \& Liu]{TrKoLi08}
Tromp, J., Komatitsch, D., \& Liu, Q., 2008.
\newblock Spectral-element and adjoint methods in seismology, {\it
  Communications in Computational Physics\/}, {\bf 3}(1), 1--32.

\bibitem[Vai et~al.(1999)Vai, Castillo-Covarrubias, S\'anchez-Sesma,
  Komatitsch, \& Vilotte]{VaCaSaKoVi99}
Vai, R., Castillo-Covarrubias, J.~M., S\'anchez-Sesma, F.~J., Komatitsch, D.,
  \& Vilotte, J.~P., 1999.
\newblock Elastic wave propagation in an irregularly layered medium, {\it Soil
  Dynamics and Earthquake Engineering\/}, {\bf 18}(1), 11--18.

\bibitem[Varela et~al.(1993)Varela, Rosa, \& Ulrych]{VaRoUl93}
Varela, C.~L., Rosa, A. L.~R., \& Ulrych, T.~J., 1993.
\newblock Modeling of attenuation and dispersion, {\it Geophysics\/}, {\bf
  58}(8), 1167--1173.

\bibitem[Virieux \& Operto(2009)]{ViOp09}
Virieux, J. \& Operto, S., 2009.
\newblock An overview of full-waveform inversion in exploration geophysics,
  {\it Geophysics\/}, {\bf 74}(6), WCC1--WCC26.

\bibitem[Wang et~al.(2009)Wang, Moin, \& Iaccarino]{WaMoIa09}
Wang, Q., Moin, P., \& Iaccarino, G., 2009.
\newblock Minimal repetition dynamic checkpointing algorithm for unsteady
  adjoint calculation, {\it SIAM Journal on Scientific Computing\/}, {\bf
  31}(4), 2549--2567.

\bibitem[Yuan et~al.(2014)Yuan, Wen, Li, Zhang, \& Liu]{YuWeLiZhLi14}
Yuan, S., Wen, S., Li, H., Zhang, X., \& Liu, Q., 2014.
\newblock An optimization framework for adjoint-based climate simulations: {A}
  case study of the {Z}ebiak-{C}ane model, {\it International Journal of High
  Performance Computing Applications\/}, {\bf 28}(2), 174--182.

\bibitem[Zhou et~al.(2011)Zhou, Liu, \& Tromp]{ZhLiTr11}
Zhou, Y., Liu, Q., \& Tromp, J., 2011.
\newblock Surface wave sensitivity: mode summation versus adjoint {SEM}, {\it
  Geophys. J. Int.\/}, {\bf 187}(3), 1560--1576.

\bibitem[Zhu \& Tromp(2013)]{ZhTr13}
Zhu, H. \& Tromp, J., 2013.
\newblock Mapping tectonic deformation in the crust and upper mantle beneath
  {E}urope and the {N}orth {A}tlantic ocean, {\it Science\/}, {\bf 341}(6148),
  871--875.

\bibitem[Zhu et~al.(2012)Zhu, Bozda{\u{g}}, Peter, \& Tromp]{ZhBoPeTr12}
Zhu, H., Bozda{\u{g}}, E., Peter, D., \& Tromp, J., 2012.
\newblock Structure of the {E}uropean upper mantle revealed by adjoint
  tomography, {\it Nature Geoscience\/}, {\bf 5}(7), 493--498.

\bibitem[Zhu(2014)]{Zhu14}
Zhu, T., 2014.
\newblock Time-reverse modelling of acoustic wave propagation in attenuating
  media, {\it Geophys. J. Int.\/}, {\bf 197}(1), 483--494.

\bibitem[Zhu et~al.(2014)Zhu, Harris, \& Biondi]{ZhHaBi14}
Zhu, T., Harris, J.~M., \& Biondi, B., 2014.
\newblock {$Q$}-compensated reverse-time migration, {\it Geophysics\/}, {\bf
  79}(3), S77--S87.

\end{thebibliography}
\end{document}